\documentclass[aps,prl,twocolumn,superscriptaddress,floatfix]{revtex4}
\usepackage[dvips]{graphicx}
\usepackage{dcolumn}
\usepackage{bm}
\usepackage{amsmath}

\begin{document}

\title{Single quasiparticle excitation dynamics on a superconducting island}

\author{V. F. Maisi}
\email{ville.maisi@mikes.fi}
\affiliation{Low Temperature Laboratory (OVLL), Aalto University School of Science, P.O.~Box 13500, 00076 Aalto, Finland}
\affiliation{Centre for Metrology and Accreditation (MIKES), P.O. Box 9, 02151 Espoo, Finland}

\author{S. V.~Lotkhov}
\affiliation{Physikalisch-Technische Bundesanstalt, Bundesallee 100, 38116 Braunschweig, Germany}

\author{A. Kemppinen}
\affiliation{Centre for Metrology and Accreditation (MIKES), P.O. Box 9, 02151 Espoo, Finland}

\author{A. Heimes}
\affiliation{Institut f\"ur Theoretische Festk\"orperphysik, Karlsruher Institut f\"ur Technologie, Wolfgang-Gaede-Str. 1, D-76128 Karlsruhe, Germany}

\author{J. T. Muhonen}
\affiliation{Low Temperature Laboratory (OVLL), Aalto University School of Science, P.O.~Box 13500, 00076 Aalto, Finland}
\affiliation{Centre for Quantum Computation and Communication Technology, School of Electrical Engineering and Telecommunications, University of New South Wales, Sydney NSW 2052, Australia}

\author{J. P. Pekola}
\affiliation{Low Temperature Laboratory (OVLL), Aalto University School of Science, P.O.~Box 13500, 00076 Aalto, Finland}

\begin{abstract} We investigate single quasiparticle excitation dynamics on a small superconducting aluminum island connected to normal metallic leads by tunnel junctions. We find the island to be free of excitations within the measurement resolution allowing us to determine Cooper pair breaking rate to be less than $3\ \mathrm{kHz}$. By tuning the Coulomb energy of the island to have an odd number of electrons, one of them remains unpaired. We detect it by measuring its relaxation rate via tunneling. By injecting electrons with a periodic gate voltage, we probe electron-phonon interaction and relaxation down to a single quasiparticle excitation pair, with a measured recombination rate of $8\ \mathrm{kHz}$. Our experiment yields a strong test of BCS-theory in aluminum as the results are consistent with it without free parameters.
\end{abstract}

\maketitle

The quasiparticle excitations describing the microscopic degrees of freedom in superconductors freeze out at low temperatures, provided no energy exceeding the superconducting gap $\Delta$ is available. Early experiments on these excitations were performed typically close to the critical temperature with large structures so that $N_S$, the number of quasiparticle excitations, was high~\cite{clarke1972,kaplan1976, parker1972,wyatt1966,dayem1967,klapwijk1976,chi1979,gray1971,wilson2001}. Later on, as the fabrication techniques progressed, it became possible to bring $N_S$ close to unity to reveal the parity effect of electrons on a superconducting island~\cite{lafarge1993,eiles1993,hergenrother1994,averin1990a, schon1994}. In recent years, the tunneling and relaxation dynamics of quasiparticles, which we address in this letter, have become a topical subject because of their influence on practically all superconducting circuits  in the low temperature limit~\cite{sun2012,leander2011,knowles2012,devisser2011,aumentado2004,ferguson2006,leppakangas2012,riste2012}.

\begin{figure}[t]
	\centering
	\includegraphics[width=0.5\textwidth]{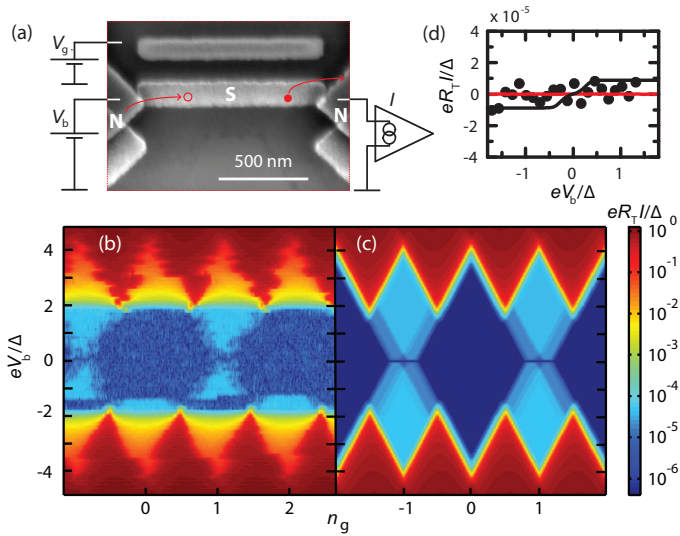}
	\caption{\label{fig:sample} (color online). (a) Scanning electron micrograph of the sample studied. It is biased with voltage $V_b$ and a gate offset voltage $V_g$ is applied to a gate electrode. The latter is not shown in the micrograph. The operation is governed by few quasiparticle excitations which can be either particle or hole like. A particle like excitation can relax by tunneling out (filled circle) and a hole by electron tunneling in and filling the state. (b) Measured source-drain current $I$ as a function of bias and gate voltages. (c) Calculated current based on sequential single-electron tunneling model. (d) Measured current at $n_g = 0$ is shown as black dots. Black line is calculated assuming a quasiparticle generation rate of $3\ \mathrm{kHz}$, and red one assuming a vanishing generation rate.}
\end{figure}

We study the quasiparticle excitations in a small aluminum island shown in Fig.~\ref{fig:sample} (a). The island is connected via a thin insulating aluminum oxide layer to two normal metallic copper leads to form a single-electron transistor (SET) allowing quasiparticle tunneling. We bias the SET by a voltage $V_b$ between the source and drain and polarize the island by an offset charge $n_g$ expressed in units of $e$ with a gate voltage $V_g\simeq en_g/C_g$. Here $C_g$ is the gate-island capacitance. 
The current $I$ through the SET is governed by sequential tunneling of single quasiparticles and it exhibits Coulomb diamonds which overlap each other because of the superconducting energy gap~\cite{pekola2008,averin2008}, observed for our structure as a region bounded by the red sawtooths in Fig.~\ref{fig:sample} (b).

In the sub-gap regime, $eV_b<2 \Delta$, of Fig.~\ref{fig:sample} (b) the current should be suppressed if there are no quasiparticle excitations present. Nonetheless, we observe a finite current which has a period twice as long in $n_g$ as compared to the high bias region, a unique feature of a superconducting island due to Cooper pairing of electrons. Its origin is a single electron unable to pair in the condensate and hence remaining as an excitation. This parity effect has been observed in the past in similar structures~\cite{lafarge1993,eiles1993,hergenrother1994} but typically with two-electron Andreev tunneling being the main transport process. We focus on devices where Andreev current is suppressed since a high charging energy, $E_c > \Delta$, makes tunneling of two electrons energetically unfavourable compared to that of a single quasiparticle~\cite{averin2008,maisi2011,aref2011}. In this case, the transport is dominated by single-electron processes allowing simple and direct probing of the quasiparticle excitations without the interfering multi-electron tunneling.

For a quantitative descripition of the transport characteristics, we performed a numerical simulation of the device operation shown in Fig.~\ref{fig:sample} (c). To describe simultaneously the charging of the island with electrons and the excitations involved in superconducting state, we assign probability $P(N,N_S)$ for  having $N$ excess electrons and $N_S$ quasiparticle excitations on the island. The time evolution of $P(N,N_S)$ is described by a master equation
\begin{equation}
\label{eq:master}
\dot{P}(N,N_S) = \sum_{N',\, N'_S} \Gamma_{N'\rightarrow N,N'_S\rightarrow N_S}P(N',N'_S),
\end{equation}
where $\dot{P}(N,N_S)$ stands for the time derivate of $P(N,N_S)$ and $\Gamma_{N'\rightarrow N,N'_S\rightarrow N_S}$ for the transition rate from state $P(N',N'_S)$ to $P(N,N_S)$. These rates are set by electron tunneling between the island and the leads, Cooper pair breaking and recombination of quasiparticles. 

Tunneling rates are calculated by the standard first order perturbation theory so that electrons tunneling into the superconducting island to a state with energy $E > \Delta$ will increase the quasiparticle number and electron number by one,
\begin{equation}
\label{eq:rate}
\Gamma_{\hspace{-3pt}\substack{N-1 \rightarrow N,\\N_S-1 \rightarrow N_S}} = \displaystyle\frac{1}{e^2 R_T}\int_\Delta^\infty \hspace{-5pt} dE \, n_S(E)(1-f_S(E))f_N(E+\delta E),
\end{equation}
where $R_T$ is the tunneling resistance, $n_S(E)$ is the BCS density of states, $f_{S/N}(E)$ the occupation probability of state $E$ in superconductor / normal metal and $\delta E$ the energy gain from charging and biasing~\cite{pekola2008,averin2008}. If the incoming electron tunnels to the lower branch $E<-\Delta$, it removes an excitation by filling a hole. Similarly, an outgoing electron from the upper branch removes an excitation while if it tunnels out from the lower branch, an excitation is generated.

In the tunneling rates the exact quasiparticle number $N_S$ is accounted for by a nonequilibrium distribution $f_S$. In general this has a complicated form as a function of energy. However due to the fact that quasiparticles are injected close to the gap, the resulting tunneling and recombination are not sensitive to the functional form of $f_S$. Our system is also symmetric with respect to the two branches. Therefore we neglect branch imbalance and paramerize the quasiparticle number by an effectively increased temperature $T_S$ and a Fermi distribution in case of $f_S$. This gives the relation
\begin{equation} \label{eq:nts}
N_S = \sqrt{2\pi}D(E_F)V\sqrt{\Delta k_BT_S}e^{-\Delta/k_BT_S},
\end{equation}
where $D(E_F)=1.45\times10^{47}{\rm J^{-1}m^{-3}}$ is the density of states in the normal state~\cite{saira2012} and $V$ the volume of the island. For the normal metallic leads we use Fermi distribution with $T_N=60\ \mathrm{mK}$, equal to the base temperature of the cryostat. For a detailed description of all transition rates, see the supplementary information~\cite{suppmat}.

For the steady-state represented by Fig.~\ref{fig:sample} (c), we solve Eq.~(\ref{eq:master}) with $\dot{P}(N,N_S)=0$ and calculate the current as an average of the tunneling rates weighted by the probabilities $P(N,N_S)$. The parameter values of sample A: $E_c = 240\ \mathrm{\mu eV}$,  $\Delta = 210\ \mathrm{\mu eV}$ and tunneling resistances $R_{T1} = 220\ \mathrm{k\Omega}$ and $R_{T2} = 150\ \mathrm{k\Omega}$ for the two junctions were used in the simulations. They were determined from measurements in the high bias regime ($e|V_b|>2\Delta$) and hence their values are independent of the sub-gap features.

The simulation of Fig.~\ref{fig:sample} (c) reproduces the behaviour observed in the experiments. The relaxation rate of a single quasiparticle excitation via tunneling is expected to be $\Gamma_{\mathrm{qp}} \equiv \Gamma_{N+1\rightarrow N,1\rightarrow 0} = (2e^2 R_TD(E_F)V)^{-1} = 190\ \mathrm{kHz}$, where we have used the measured dimensions for $V =1.06 \,\mu {\rm m} \times 145 \,{\rm nm} \times 25 {\rm nm}$. From the fit in the sub-gap regime we obtain $\Gamma_{\mathrm{qp}} = 150\ \mathrm{kHz}$, consistent with the prediction. The value of $\Gamma_{\mathrm{qp}}$ affects only the value of current on the light blue plateau of Fig.~\ref{fig:sample} (c), not the actual form or size of the terrace. As our simulation based on sequential tunneling reproduces all the features in the sub-gap regime, we conclude that the two-electron periodicity originates from single-electron tunneling and the operation is essentially free of multi-electron tunneling processes. At odd integer values of $n_g$, the characteristic feature of Andreev tunneling would be a linear-in-$V_b$ current at low bias voltages and a subsequent drop  \cite{eiles1993,hekking1993}, which is absent in our data. The leakage current in the sub-gap region does not vanish even in the zero temperature limit but remains essentially the same as presented in Fig.~\ref{fig:sample}. Therefore, all quasiparticle excitations cannot be suppressed at finite bias voltages by lowering the temperature, if $n_g$ is close to an odd integer.

At even integer values of $n_g$ we have ideally no current flow as all electrons are paired. If Cooper pair breaking would take place in the island, we would obtain two quasiparticle excitations in the superconductor. One of these excitations can then relax by tunneling to the leads followed by tunneling of a new excitation to neutralize the offset charge, similarly as at the odd $n_g$ case described above. This cycle would continue until the two quasiparticles  recombine to a Cooper pair. As the recombination rate, discussed below, is slower than the rates in the cycle, we have several electrons tunneling through the device for each broken pair, hence amplifying the signal. With this model we obtain an upper bound  $\Gamma_{N\rightarrow N,0\rightarrow 2} = 3\ \mathrm{kHz}$ for the pair breaking rate under our experimental conditions based on simulations shown in Fig.~\ref{fig:sample} (d). This generation rate corresponds to energy absorption at $2 \Gamma_{N\rightarrow N,0\rightarrow 2} \Delta = 0.1\ \mathrm{aW}$ power on the superconducting island.

Under constant biasing conditions, there is at most one quasiparticle present in the subgap regime at low temperatures. The non-tunneling relaxation on the island is then not possible, since recombination would call for two excitations. Therefore the static case can be described by pure tunneling without other relaxation processes. To study recombination of two quasiparticles into a Cooper pair, we injected intentionally more quasiparticles to the island. The injection was done by a periodic drive of the gate voltage. By changing $n_g$, we change the potential of the superconducting  island and either pull quasiparticle excitations into the island when the potential is lowered or create hole type excitations as potential is raised and quasiparticles tunnel out. The number of injected quasiparticles and the number of quasiparticles on the island can then be determined from the resulting current curves with the help of simulations. In the experiment we approach two different limits which we will discuss in the following: When the pumping frequency is high, $N_S$ is large. Such situation can be described by a \emph{thermal model}, i.e., by an increased time-independent effective temperature of the superconducting island. In the opposite limit of low frequency, $N_S$ is small. Then the thermal model fails and we have to account for the exact time dependent number of quasiparticles.

\begin{figure}[t]
	\centering
	\includegraphics[width=0.49\textwidth]{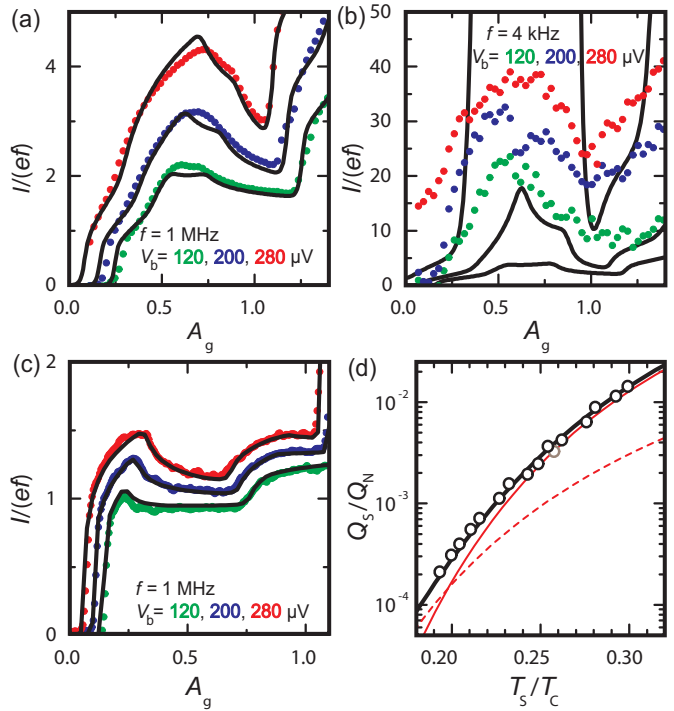}
	\caption{\label{fig:plateaus} (color online). (a)-(b) Measured current of Sample A against gate voltage amplitude $A_g$ at $f = 4$ and $1000\ \mathrm{kHz}$ at bias voltage values $V_b =120, 200, 280\ \mathrm{\mu V}$ shown as blue, red and green dots respectively. Black lines show simulations assuming an elevated temperature on the superconducting island. (c) Similar measurement for Sample B at $f =1\ \mathrm{MHz}$. (d) The electron-phonon heat flux in the superconducting state normalized by that in the normal state extracted from the measurements (black circles). Temperature is expressed with respect to the critical temperature $T_C=\Delta/ 1.76 k_B$. The theoretical result of Eq.~(\ref{eq:eph}) is shown by the black line. Solid and dotted red lines show the recombination and scattering part of Eq.~(\ref{eq:eph}) correspondinly. The open grey symbol is from Sample B.}
\end{figure}

\begin{figure}[t]
	\centering
	\includegraphics[width=0.49\textwidth]{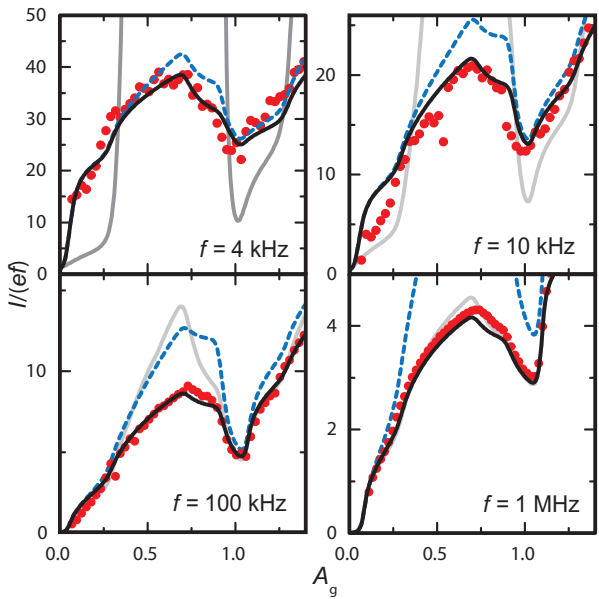}
	\caption{\label{fig:plateaus2} (color online). Red circles show the measured current for $f = 4 - 1000\ \mathrm{kHz}$ at $V_b = 280\ \mathrm{\mu V}$. Black lines are simulations based on Eq.~(\ref{eq:master}). Recombination rates are taken to match the heat flux in the regime where the thermal model applies (Eq.~(\ref{eq:eph})). Dotted blue lines are calculated with vanishing electron-phonon relaxation rate and solid gray lines with the thermal model.}
\end{figure}

In Fig.~\ref{fig:plateaus} (a) we show the measured current for three different values of bias voltage $V_b$. The gate drive is sinusoidal around $n_g = 1/2$ with amplitude $A_g$ expressed in units of $e/C_g$ and frequency $f = 1\ \mathrm{MHz}$ corresponding to fast pumping. Without accumulation of quasiparticles to the island, the current would show quantized plateaus with spacing $ef$, similar to the SINIS turnstile~\cite{pekola2008}. However, as the island has a surplus of quasiparticles, i.e. it is heated up, the current is substantially higher. We use now the thermal model where the heat injection to the island by electron tunneling is balanced by electron-phonon interaction. The heat flux into the phonon bath is given by
\begin{equation}
\label{eq:eph}
\begin{array}{ccl}
\dot{Q}_\mathrm{ep} &=& \frac{\Sigma V}{24 \zeta(5) k_B^5} \int_0^\infty d\epsilon\ \epsilon^3 \left(n(\epsilon,T_S)-n(\epsilon,T_P) \right) \int_{-\infty}^\infty dE  \\
 & & \hspace{-35pt} \times n_S(E) n_S(E+\epsilon) \left(1 - \frac{\Delta^2}{E(E+\epsilon)} \right)\left(f_S(E)-f_S(E+\epsilon) \right),
\end{array}
\end{equation}
where $\Sigma$ is the material constant for electron-phonon coupling, $\zeta(z)$ the Riemann zeta function and $n(\epsilon,T) = (\exp(\epsilon/(k_BT))-1)^{-1}$  the Bose-Einstein distribution of the phonons at temperature $T_P$. See supplementary information for derivation. The same result is obtained by kinetic Boltzmann equation calculations~\cite{averin1990,timofeev2009}. 
The simulations based on the thermal model are shown as black lines in Fig.~\ref{fig:plateaus} (a). 
The electron-phonon coupling constant $\Sigma = 1.8 \cdot 10^9\ \ \mathrm{WK^{-5}m^{-3}}$, used in simulations, was measured in the normal state, where $\dot{Q}_\mathrm{ep}=\Sigma V (T_N^5-T_P^5)$ and $T_N$ is the electron temperature~\cite{wellstood1994}.

We expect the thermal model to be a good approach if $N_S\gg 1$. With the high frequency and large amplitude drive in Fig.~\ref{fig:plateaus} (a), we have $N_S\sim 10$ quasiparticles present for $A_g \sim 1$, suggesting that the thermal model is adequate for these data. If the frequency is lowered to $f = 4\ \mathrm{kHz}$, shown in Fig.~\ref{fig:plateaus} (b), the thermal model fails as $N_S$ approaches unity. As a further proof of the overheating, we repeated the high frequency measurement using sample B with measured parameter values $E_c = 620\ \mathrm{\mu eV}$, $\Delta = 270\ \mathrm{\mu eV}$, $R_{T1} = 1800\ \mathrm{k\Omega}$, $R_{T2} = 960\ \mathrm{k\Omega}$ and $V = 800\ \mathrm{nm} \times  60\ \mathrm{nm} \times  15\ \mathrm{nm}$. The result is shown in Fig~\ref{fig:plateaus} (c). Again, the simulations (black lines) are able to reproduce all non-trivial features of the measured curves. As a summary of the thermal model fits, we repeated the measurement of Fig.~\ref{fig:plateaus} (a) at different frequencies and determined by numerical simulations the temperature of the superconducting island and the heat injected into it based on the measured current. The results are shown in Fig.~\ref{fig:plateaus} (d) as black circles. The results match well with the expected electron-phonon coupling of the superconductor, Eq.~(\ref{eq:eph}), presented as the solid black line.

For studying the deviation from the thermal model, we measured the characteristics at four frequencies $f =4, 10, 100$ and $1000\ \mathrm{kHz}$ at $V_b = 280\ \mathrm{\mu V}$, shown in Fig.~\ref{fig:plateaus2}. The thermal model is presented now as solid gray lines. A more adequate description of the data at low frequencies is obtained by simulations based on Eq.~(\ref{eq:master}), where we keep track on the number of excitations during the operation and take into account the recombination rates. The heat flux of Eq.~(\ref{eq:eph}) consists of recombination and scattering which can, respectively, be expressed as
\begin{equation}
  \label{eq:sc_rec}
\begin{array}{ccl}
  \dot Q_\mathrm{rec} &=& \frac{\pi V\Sigma}{3 \zeta(5) k_B^5}\left(k_BT_S \Delta^4+\frac{7}{4}(k_BT_S)^2 \Delta^3 \right) e^{-2\Delta/k_BT_S}, \\
  \dot Q_\mathrm{sc}  &=& {V\Sigma}T_S^5 e^{-\Delta/k_BT_S}.
\end{array}
\end{equation}
Their contributions are presented in Fig.~\ref{fig:plateaus} (d). Whereas the scattering does not change the number of quasiparticles $N_S$, recombination leads to transitions $N_S\rightarrow N_S-2$. We account for this process by including the recombination rate $\Gamma_{N\rightarrow N,N_S\rightarrow N_S-2}={\dot Q_\mathrm{rec}(N_S)}/{2\Delta}$, where the relation between the effective temperature $T_S$ in Eq.~\eqref{eq:sc_rec} and the exact quasiparticle number $N_S$ is given by Eq.~\eqref{eq:nts}. The solid black lines of Fig.~\ref{fig:plateaus2} are calculated with the same value of $\Sigma$ as obtained in the normal state. Blue dotted lines show similar simulations where electron-phonon relaxation is disregarded. 

With the simulations based on instantaneous quasiparticle number we can reproduce the experimental features precisely with no free parameters in the calculation. At the lowest frequency, $f = 4\ \mathrm{kHz}$, we have only one quasiparticle present for most of the time. Hence the curves are not sensitive to the recombination. As frequency is increased, the simulations without electron-phonon relaxation deviate from the experimental data. For $f = 10\ \mathrm{kHz}$ we probe the recombination rate of a single qp pair only, $\Gamma_{N\rightarrow N,2\rightarrow 0} = 8\ \mathrm{kHz}$. We checked this by artificially changing the recombination rates for $N_S>2$, without any significant difference in the curves. At higher frequencies, the recombination for $N_S > 2$ becomes significant as well. The results of the two models approach each other and the thermal model becomes valid.

In summary, a small superconducting island at low temperatures has allowed us to study the dynamics of single electronic excitations and their relaxation. Under quiescent conditions we found a vanishing Cooper pair breaking rate within the measurement resolution: based on the measurement noise we obtained an upper limit of $3\ \mathrm{kHz}$ for this rate. On the other hand, by periodically pumping electrons, we controllably increased the number of quasiparticles and were able to measure the recombination rates both in the large quasiparticle number limit and for a single quasiparticle pair: $\Gamma_{N\rightarrow N,2\rightarrow 0} = 8\ \mathrm{kHz}$. The recombination rates are in quantitative agreement with the relaxation measured at higher temperatures and in the normal state.

We thank D. V. Averin, G. Sch\"on, F. W. J. Hekking, D. Golubev, T. Heikkil\"a and A. Zorin for discussions. The work has been supported partially by LTQ (project no. 250280) CoE grant, the European Community's Seventh Framework Programme under Grant Agreement No. 238345 (GEOMDISS) and the National Doctoral Programme in Nanoscience (NGS-NANO). We acknowledge the provision of facilities and technical support by Aalto University at Micronova Nanofabrication Centre.

\end{document}


\title{Single quasiparticle excitation dynamics on a superconducting island \linebreak -\linebreak   Supplementary information}

\author{V. F. Maisi}
\email{ville.maisi@mikes.fi}
\affiliation{Low Temperature Laboratory (OVLL), Aalto University School of Science, P.O.~Box 13500, 00076 Aalto, Finland}
\affiliation{Centre for Metrology and Accreditation (MIKES), P.O. Box 9, 02151 Espoo, Finland}

\author{S. V.~Lotkhov}
\affiliation{Physikalisch-Technische Bundesanstalt, Bundesallee 100, 38116 Braunschweig, Germany}

\author{A. Kemppinen}
\affiliation{Centre for Metrology and Accreditation (MIKES), P.O. Box 9, 02151 Espoo, Finland}

\author{A. Heimes}
\affiliation{Institut f\"ur Theoretische Festk\"orperphysik, Karlsruher Institut f\"ur Technologie, Wolfgang-Gaede-Str. 1, D-76128 Karlsruhe, Germany}

\author{J. T. Muhonen}
\affiliation{Low Temperature Laboratory (OVLL), Aalto University School of Science, P.O.~Box 13500, 00076 Aalto, Finland}
\affiliation{Centre for Quantum Computation and Communication Technology, School of Electrical Engineering and Telecommunications, University of New South Wales, Sydney NSW 2052, Australia}

\author{J. P. Pekola}
\affiliation{Low Temperature Laboratory (OVLL), Aalto University School of Science, P.O.~Box 13500, 00076 Aalto, Finland}

\maketitle

{\bf MASTER EQUATION} \\
We model the superconducting island of the main article by writing down a master equation by assigning a probability $P(N,N_S)$ for having $N$ excess electrons and $N_S$ quasiparticles on the island. The time derivate of $P(N,N_S)$ is then 
\begin{equation}
\label{eq:master}
\dot{P}(N,N_S) = \sum_{N',\, N'_S} \Gamma_{N'\rightarrow N,N'_S\rightarrow N_S}P(N',N'_S),
\end{equation}
where  $\Gamma_{N'\rightarrow N,N'_S\rightarrow N_S}$ is a transition rate from state $(N',N'_S)$ to $(N,N_S)$.  For $N' = N, N'_S = N_S$ we insert a rate $\Gamma_{N\rightarrow N,N_S\rightarrow N_S} = - \sum_{N',\, N'_S} \Gamma_{N\rightarrow N'\neq N,N_S\rightarrow N'_S\neq N_S}$, which is a sum of all rates out from the state $(N,N_S)$. Next we consider the rates arising from electron tunneling between normal metallic leads and the island and the recombination and pair breaking in the superconductor.
\vspace{5 mm}

\begin{figure}[t]
	\centering
	\includegraphics[width=0.65\textwidth]{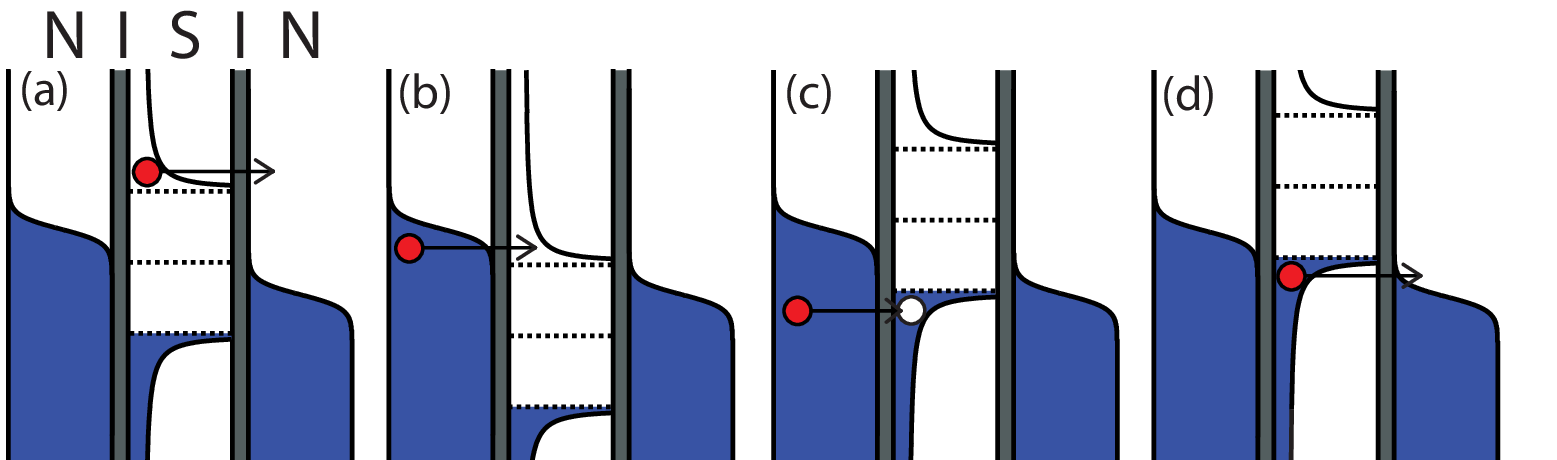}
	\caption{\label{fig:dos} (color online). The four different tunneling processes changing $N$ and $N_S$. (a) Electron tunneling out of the island and removing an excitation. (b) Electron tunneling into the island creating and excitation. (c) Electron tunneling into the island and removing a hole excitation. (d) Electron tunneling out and creating a hole excitation.}
\end{figure}

{\bf ELECTRON TUNNELING} \\
Within first order, the tunneling rates are given by Fermi golden rule similarly as in Refs.~\cite{averin1991}. Now as we consider also explicitly $N_S$, we need to split the rates so that an electron tunneling into the upper branch of the superconducting island to energy $E > \Delta$, increases the excitation number $N_S$ by one and an electron tunneling into lower branch, $E < -\Delta$, decrease it by one as shown in Fig.~\ref{fig:dos}. Likewise, an electron tunneling out from the upper branch decreases $N_S$ and an electron tunneling out from the lower branch increases it by one. Hence we obtain rates
\begin{equation}
\left\{
\begin{array}{ccl}
\Gamma_{N-1\rightarrow N,\, N_S-1\rightarrow N_S} &=&  \displaystyle \sum_{j=1,2}\frac{1}{e^2R_{\mathrm{T},j}}\int_\Delta^\infty dE\ n_S(E) (1-f_S(E,N_S))f_N(E+\delta E_j(N))\vspace{5pt}  \\
\Gamma_{N-1\rightarrow N,\, N_S+1\rightarrow N_S}  &=&  \displaystyle  \sum_{j=1,2} \frac{1}{e^2R_{\mathrm{T},j}}\int_{-\infty}^{-\Delta} dE\ n_S(E) (1-f_S(E,N_S))f_N(E+\delta E_j(N))\vspace{5pt} \\
\Gamma_{N+1\rightarrow N,\, N_S-1\rightarrow N_S} &=& \displaystyle   \sum_{j=1,2} \frac{1}{e^2R_{\mathrm{T},j}}\int_{-\infty}^{-\Delta} dE\ n_S(E) f_S(E,N_S)(1-f_N(E+\delta E_j(N)))\vspace{5pt}\\
\Gamma_{N+1\rightarrow N,\, N_S+1\rightarrow N_S} &=& \displaystyle   \sum_{j=1,2} \frac{1}{e^2R_{\mathrm{T},j}}\int_\Delta^\infty dE\ n_S(E) f_S(E,N_S)(1-f_N(E+\delta E_j(N))),  
\end{array}
\right.
\end{equation}
where $R_{\mathrm{T},j}$ is the tunneling resistance of junction $j$, $n_S(E)$, the BCS density of states, $f_S(E,N_S)$, occupation probability for state $E$ on the island if we have $N_S$ quasiparticles. and $f_N(E)$ occupation probability in the normal metallic lead at energy $E$. $\delta E_j(N)$ the energy cost from biasing and charging energy~\cite{averin1991}. We assume $f_N$ to be a Fermi distribution with temperature equal to the bath temperature $T_N = 60\ \mathrm{mK}$. For the superconductor we need to have $f_S$ such that the number of excitations equals $N_S$. For the tunneling rates, as well as for the electron-phonon interaction discussed below, the functional form of the distribution is irrelevant as long as the excitations are close to the gap edges $E \approx \pm \Delta$, where we inject them. We also neglect the branch imbalance, i.e. assume no chemical potential shift, $\delta \mu = 0$, because our drive is symmetrical with respect to the two branches. Hence we use a Fermi distribution with temperature $T_S$. The relation between $N_S$ and $T_S$ is then
\begin{equation}
\label{eq:ns}
 N_S= 2 D(E_F) V \int_\Delta^\infty dE \frac{E}{\sqrt{E^2-\Delta^2}}f_E\approx\sqrt{2\pi} D(E_F) V \sqrt{\Delta k_B T_S}e^{-\Delta/k_BT_S},
\end{equation}
where $D(E_F)$ is the normal state density of states at Fermi level and $V$ the volume of the island. The tunneling current is obtained then as
\begin{equation}
\label{eq:current}
I = e\sum_{N, N_S, \pm}\Big( \Gamma_{N\rightarrow N+1,\, N_S\rightarrow N_S\pm 1} 
 -\Gamma_{N\rightarrow N-1,\, N_S \rightarrow N_S \pm 1} \Big)P\left( N,N_S\right).
\end{equation}
Including higher order tunneling to Eqs.~(\ref{eq:master}) and (\ref{eq:current}) is straightforward. For example Andreev current gives rise to terms $\Gamma_{N \pm 2\rightarrow N,\, N_S \rightarrow N_S}$. These rates are calculated similarly as in Refs.~\cite{averin2008,maisi2011,aref2011} since quasiparticle number $N_S$ is not changing. Including Andreev tunneling terms into the calculations in main text, does not change the results as $E_c > \Delta$. For $E_c < \Delta$, those terms contribute because Andreev tunneling thresholds are exceeded before the single-electron tunneling thresholds~\cite{averin2008,maisi2011,aref2011}. 

\vspace{5 mm}

{\bf ELECTRON-PHONON INTERACTION} \\
To obtain the recombination rates for Eq.~(\ref{eq:master}) and Eq.~(4) of the main article we use Hamiltonian
\begin{equation}
H = H_s + H_p + H_\mathrm{ep}.
\end{equation}
Here $ H_s + H_p$ is the non-perturbated part and $H_\mathrm{ep}$ the perturbation from electron-phonon interaction. The BCS Hamiltonian of the superconductor is
\begin{equation}
H_s = \sum_{k\sigma} E_k \gamma_{k\sigma}^\dagger \gamma_{k\sigma},
\end{equation}
with $E_k = \sqrt{\epsilon_k^2+|\Delta_k|^2}$. The diagonalizing fermionic operators $\gamma_{k\sigma}$ are connected to the electron operators $c_{k'\sigma'}$ by $c_{k\sigma}^\dagger  = v_{k\sigma}^* \gamma_{-(k\sigma)} + u_{k\sigma}^{} \gamma_{k\sigma}^\dagger$,
where we use notation
\begin{equation}
\begin{array}{cc}
v_{k\sigma} = 
\left\{
\begin{array}{ll}
v_k & \text{if } \sigma = \uparrow \\
-v_{-k} & \text{if }  \sigma = \downarrow 
\end{array}
\right.
& \hspace{30pt}
u_{k\sigma} = 
\left\{
\begin{array}{ll}
u_k & \text{if } \sigma = \uparrow \\
u_{-k} & \text{if }  \sigma = \downarrow 
\end{array}
\right. .
\end{array}
\end{equation}
The coefficients satisfy $|u_k|^2 + |v_k|^2 = 1$ and $\Delta_k^*v_k/u_k = E_k - \epsilon_k$ leading to 
\begin{equation}
|v_k|^2 = 1 - |u_k|^2 = \frac{1}{2}\left(1 - \frac{\epsilon_k}{E_k} \right).
\end{equation}
Similarly, we have for the phonons
\begin{equation}
H_p = \sum_q \hbar \omega_q b_q^\dagger b_q.
\end{equation}
The coupling of these two systems has the form
\begin{equation}
H_\mathrm{ep} = \nu \sum_{k,q} \omega_q^{1/2} ( c_k^\dagger c_{k-q} b_q + c_k^\dagger c_{k+q} b_q^\dagger ).
\end{equation}
as in Ref.~\cite{fw}. The operator of heat flux to the phonons is
\begin{equation}
\label{eq:heatFluxDef}
\dot{H}_{p} = \frac{i}{\hbar}[H,H_p] = \frac{i}{\hbar}[H_\mathrm{ep},H_p] = i \nu \sum_{k,q} \omega_q^{3/2} ( c_k^\dagger c_{k-q} b_q - c_k^\dagger c_{k+q} b_q^\dagger ).
\end{equation}
By using the Kubo formula in the interaction picture we obtain
\begin{equation}
\label{eq:heatFluxDef}
\begin{array}{ccl}
\left<\dot{H}_{p}\right> &=& \left<\dot{H}_{p}\right>_0 - \displaystyle\frac{i}{\hbar}\displaystyle\int_{-\infty}^t dt'\left< \left[\dot{H}_p(t),H_\mathrm{ep}(t')\right]\right>_0 \\
 & = & \displaystyle\frac{1}{\hbar} \sum_{\substack{k\sigma q \\ k'\sigma'}} \omega_q^2 \displaystyle\int_{-\infty}^t dt' \Bigg\{  \Bigg.
\vspace{5pt} \\
 &    &  \left< \left[c_{k\sigma}^\dagger(t)\: c_{k-q\sigma}(t)\: b_q(t),\: c_{k'\sigma'}^\dagger(t')\: c_{k'+q\sigma'}(t')\: b_q^\dagger(t')\right]\right>_0 - \\
 &    &  \left< \left[c_{k\sigma}^\dagger(t)\: c_{k+q\sigma}(t)\: b_q^\dagger(t),\: c_{k'\sigma'}^\dagger(t')\: c_{k'-q\sigma'}(t')\: b_q(t')\right]\right>_0 
 \Bigg. \Bigg\}.
\end{array}
\end{equation}
Next we use $b_q(t) =  e^{-i\omega_q t}  b_q$ and $c_{k\sigma}^\dagger(t)  = v_{k\sigma}^* \gamma_{-(k\sigma)}(t) + u_{k\sigma}^{} \gamma_{k\sigma}^\dagger(t)$ with $\gamma_{k\sigma}(t) =  \gamma_{k\sigma} e^{-i E_k t / \hbar}$. As an example, let us consider the first non-zero term in Eq.~(\ref{eq:heatFluxDef}). We obtain
\begin{equation}
\begin{array}{ccl}
& & \displaystyle e^{i \omega_q(t'-t)} e^{i(-E_k t + E_{k-q} t - E_{k'} t' + E_{k'+q} t')/\hbar} v_{k\sigma}^* v_{k-q\sigma}v_{k'\sigma'}^*v_{k'+q\sigma'} \left< [\gamma_{-(k\sigma)}^{} \gamma_{-(k-q\sigma)}^\dagger b_q , \gamma_{-(k'\sigma')}^{} \gamma_{-(k'+q\sigma')}^\dagger b_q^\dagger] \right>_0 \\
&=&\displaystyle  e^{i \omega_q(t'-t)} e^{i(E_{k-q}-E_k)(t- t')/\hbar} |v_{k\sigma}|^2 |v_{k-q\sigma}|^2 \left< [\gamma_{-(k\sigma)}^{} \gamma_{-(k-q\sigma)}^\dagger b_q , \gamma_{-(k-q\sigma)}^{} \gamma_{-(k\sigma)}^\dagger b_q^\dagger] \right>_0 \\
&=&\displaystyle e^{i \omega_q(t'-t)} e^{i(E_{k-q}-E_k)(t- t')/\hbar} |v_{k\sigma}|^2 |v_{k-q\sigma}|^2\Big\{
\left(1-f_k\right)f_{k-q}\left(n_q+1\right) - f_k\left(1-f_{k-q}\right)n_q
\Big\},
\end{array}
\end{equation}
where we have paired the operators as $k = k'+q$. For some of the terms, there exists two different pairing possibilities. From the second commutator of Eq.~(\ref{eq:heatFluxDef}) we find similarly a term
\begin{equation}
\begin{array}{ccl}
& &\displaystyle e^{-i \omega_q(t'-t)} e^{-i(E_{k-q}-E_k)(t- t')/\hbar} |v_{k\sigma}|^2 |v_{k-q\sigma}|^2\Big\{
\left(1-f_k\right)f_{k-q}\left(n_q+1\right) - f_k\left(1-f_{k-q}\right)n_q
\Big\},
\end{array}
\end{equation}
where the dummy summation index has been changed from $k'$ to $k$. Now we can combine these two terms and do the time integration to obtain energy conservation rules as delta functions. By combining all terms similarly, assuming $\Delta_k = \Delta e^{i\phi}$, where $\Delta$ is real, and changing the summing to integration we obtain from Eq.~(\ref{eq:heatFluxDef})
\begin{equation}
\begin{array}{l}
\dot{Q}_\mathrm{ep} = \pi \nu^2 N(E_F) D(q) \displaystyle \int_{-\infty}^\infty d \epsilon_k \int d^3q \; \omega_q^2 \Big\{ \Big. \vspace{2pt} \\
\left(1-\frac{\Delta^2}{E_k E_{k-q}}\right)
      \Big(\left(1-f_k\right)f_{k-q}\left(n_q+1\right) - f_k\left(1-f_{k-q}\right)n_q \Big) \delta(E_k-E_{k-q}+\hbar\omega_q) + \\
\left(1+\frac{\Delta^2}{E_k E_{k-q}}\right)
      \Big(\left(1-f_k\right)\left(1-f_{k-q}\right) \left(n_q+1\right) - f_k f_{k-q} n_q \Big)\delta(E_k+E_{k-q}+\hbar\omega_q)  + \\
\left(1+\frac{\Delta^2}{E_k E_{k-q}}\right)
      \Big(f_k f_{k-q}\left(n_q+1\right) - \left(1- f_k\right)\left(1-f_{k-q}\right)n_q \Big) \delta(-E_k-E_{k-q}+\hbar\omega_q) + \\
\left(1-\frac{\Delta^2}{E_k E_{k-q}}\right)
      \Big(f_k\left(1-f_{k-q}\right)\left(n_q+1\right) - \left(1-f_k\right)f_{k-q}n_q \Big)\delta(-E_k+E_{k-q}+\hbar\omega_q)  \Big\},
\end{array}
\end{equation}
where $N(E_F)$ is the density of states of the normal state and $D(q) = \frac{V}{(2\pi)^3}$ the density of states of the phonons. $V$ is the volume of the system. Now, $\int d^3q = 2 \pi \int_0^\infty dq q^2 \int_{-1}^1 d(\cos(\theta))$ where $\theta$ is the angle between $k$ and $q$. Further more $\epsilon_k = \frac{\hbar^2 k^2}{2m}$, $\epsilon_{k-q} = \frac{\hbar^2 (k-q)^2}{2m} = \epsilon_k \frac{\hbar^2 k_F}{m}q \cos(\theta)$ and $\omega_q = c_l q$, where $c_l$ is the speed of sound. For integrating the delta functions over $\cos(\theta)$, we need
\begin{equation}
\left| \frac{d E_{k-q}}{d(\cos(\theta))} \right|^{-1} = \frac{E_{k-q}}{\epsilon_{k-q}} \frac{m}{\hbar^2 k_F q} = n_S(E_{k-q})\: \frac{m}{\hbar^2 k_F q},
\end{equation}
where we have identified the density of states of the superconductor $n_S(E) = \frac{E}{\epsilon} = \frac{E}{\sqrt{E^2-\Delta^2}}$. Therefore we obtain
\begin{equation}
\label{eq:interheat}
\begin{array}{rl}
\dot{Q}_\mathrm{ep} =& \displaystyle \frac{\nu^2 V}{2\pi} N(E_F) c_l^2  \int_{-\infty}^\infty d \epsilon_k \int_0^\infty dq\: q^3 \frac{m}{\hbar^2 k_F} n_S(E_{k-q}) \Big\{ \Big. \vspace{2pt} \\
&\left(1-\frac{\Delta^2}{E_k E_{k-q}}\right)
      \Big(\left(1-f_k\right)f_{k-q}\left(n_q+1\right) - f_k\left(1-f_{k-q}\right)n_q \Big) \Big|_{E_{k-q} = E_k+\hbar\omega_q > 0} \\
+&\left(1+\frac{\Delta^2}{E_k E_{k-q}}\right)
      \Big(\left(1-f_k\right)\left(1-f_{k-q}\right) \left(n_q+1\right) - f_k f_{k-q} n_q \Big) \Big|_{E_{k-q} = -E_k-\hbar\omega_q > 0} \\
+&\left(1+\frac{\Delta^2}{E_k E_{k-q}}\right)
      \Big(f_k f_{k-q}\left(n_q+1\right) - \left(1- f_k\right)\left(1-f_{k-q}\right)n_q \Big) \Big|_{E_{k-q} = -E_k+\hbar\omega_q > 0} \\
+&\left(1-\frac{\Delta^2}{E_k E_{k-q}}\right)
      \Big(f_k\left(1-f_{k-q}\right)\left(n_q+1\right) - \left(1-f_k\right)f_{k-q}n_q \Big) \Big|_{E_{k-q} = E_k-\hbar\omega_q > 0}\ \  \Big\}.
\end{array}
\end{equation}
Now, we change the integration variables to $E_k = \sqrt{\epsilon_k^2+\Delta^2}$ and $\epsilon = \hbar \omega_q = \hbar c_l q$. Hence, we have $d\epsilon_k = n_S(E_k)dE_k$ and $dq = (\hbar c_l)^{-1} d\epsilon$. We can also use notation $E = E_k$ and $E' = E_{k-q}$. Then, equation~(\ref{eq:interheat}) yields
\begin{equation}
\label{eq:interheat2}
\begin{array}{rl}
\dot{Q}_\mathrm{ep} =& \displaystyle \frac{m \nu^2 V N(E_F)}{\pi \hbar^6 k_F c_l^2}  \int_{0}^\infty dE \int_0^\infty d\epsilon\: \epsilon^3  n_S(E) n_S(E') \Big\{ \Big. \vspace{2pt} \\
&\left(1-\frac{\Delta^2}{E E'}\right)
      \Big(\left(1-f(E)\right)f(E')\left(n(\epsilon)+1\right) - f(E)\left(1-f(E')\right)n(\epsilon) \Big) \Big|_{E' = E+\epsilon > 0} \\
+&\left(1+\frac{\Delta^2}{E E'}\right)
      \Big(\left(1-f(E)\right)\left(1-f(E')\right) \left(n(\epsilon)+1\right) - f(E) f(E') n(\epsilon) \Big) \Big|_{E' = -E-\epsilon > 0} \\
+&\left(1+\frac{\Delta^2}{E E'}\right)
      \Big(f(E) f(E')\left(n(\epsilon)+1\right) - \left(1- f(E)\right)\left(1-f(E')\right)n(\epsilon) \Big) \Big|_{E' = -E+\epsilon > 0} \\
+&\left(1-\frac{\Delta^2}{E E'}\right)
      \Big(f(E)\left(1-f(E')\right)\left(n(\epsilon)+1\right) - \left(1-f(E)\right)f(E')n(\epsilon) \Big) \Big|_{E' = E-\epsilon > 0}\ \  \Big\}.
\end{array}
\end{equation}
Finally we can change integration variable $E$ and the dummy variable $E'$ as follows: for first term: no changes, for second term: change of sign for $E'$, for third term: change of sign for $E$ and for the fourth term: change of sign for $E$ and $E'$. Then, we see that both $E$ and $E'$ cover both positive and negative values and we obtain
\begin{equation}
\label{eq:interheat3}
\begin{array}{rl}
\dot{Q}_\mathrm{ep} =& \displaystyle \frac{m \nu^2 V N(E_F)}{\pi \hbar^6 k_F c_l^3}  \int_{-\infty}^\infty dE \int_0^\infty d\epsilon\: \epsilon^3  n_S(E) n_S(E+\epsilon) \vspace{5pt} \times \\
&\left(1-\frac{\Delta^2}{E(E+\epsilon)}\right)
      \Big(\left(1-f\left(E\right)\right)f\left(E+\epsilon\right)\left(n\left(\epsilon\right)+1\right) - f\left(E\right)\left(1-f\left(E+\epsilon\right)\right)n\left(\epsilon\right) \Big).
\end{array}
\end{equation}
With some algebra, we can simplify Eq.~(\ref{eq:interheat}) to
\begin{equation}
\label{eq:interheat4}
\dot{Q}_\mathrm{ep} = \frac{\Sigma V}{24 \zeta(5) k_B^5} \int_0^\infty d\epsilon\ \epsilon^3 \left(n(\epsilon,T_S)-n(\epsilon,T_P) \right) \int_{-\infty}^\infty dE \  n_S(E) n_S(E+\epsilon) \left(1 - \frac{\Delta^2}{E(E+\epsilon)} \right)\left(f(E)-f(E+\epsilon) \right).
\end{equation}
Here we have identified the electron-phonon coupling constant $\Sigma = \frac{24 \zeta(5) k_B^5 m \nu^2 N(E_F) }{\pi \hbar^6 k_F c_l^3}$ by taking the limit $\Delta \rightarrow 0$, which yields the normal-state result $\dot{Q}_\mathrm{ep}=\Sigma V (T_S^5-T_P^5)$. One arrives to Eq.~(\ref{eq:interheat4}) also by using kinetic equations for the electronic excitations under electron-phonon interaction~\cite{Baryakhtar1979}.

For obtaining the recombination and scattering rates, we identify in Eq.~(\ref{eq:interheat2}) the first and last term to correspond to scattering and the two middle ones to recombination and pair breaking. Since the heat rate is set dominantly by the quasiparticle occupation near the gap, we can again parametrize the number of quasiparticles by an effective temperature $T_S$ and the Fermi distribution $f_E=(\exp\left({E}/{k_B T_S}\right)+1)^{-1}$ without any loss off generality. For $T_P \ll T_S \ll \Delta/k_B$, satisfied in the experiments, we obtain the recombination and scattering heat fluxes as 
\begin{eqnarray*}
  \dot Q_\mathrm{rec}(T_S) &\approx& \frac{\pi b}{2\hbar} D(E_F)V\int_{-\infty}^\infty d\xi\int_{-\infty}^\infty d\xi' (E+E')^3  \left[1+\frac{\Delta^2}{EE'}\right]f_{E} f_{E'}  \\
  &\approx& \frac{\pi V\Sigma}{3 \zeta(5) k_B^5}(k_BT_S \Delta^4+\frac{7}{4}(k_BT_S)^2 \Delta^3) e^{-2\Delta/k_BT_S},\\
  \dot Q_\mathrm{sc}(T_S)&\approx&  \frac{\pi b}{\hbar} D(E_F)V\int_{-\infty}^\infty d\xi\int_{-\infty}^\infty d\xi' \theta(E-E') (E-E')^3\left[1-\frac{\Delta^2}{EE'}\right] f_{E} \\
  &\approx& {V\Sigma}T_S^5 e^{-\Delta/k_BT_S}.
\end{eqnarray*}
Instead of parametrisation with respect to $T_S$, we could also parametrize the effective temperature $T_S$ by the quasiparticle number $N_S$ with Eq.~(\ref{eq:ns}) and ask for the rate $\Gamma_{N\rightarrow N,N_S+2\rightarrow N_S}$ entering the master equation for $P(N,N_S)$. This accounts for the transition of $N_S+2$ to $N_S$ quasiparticles, whereas the number of excess electrons $N$ is unchanged. As the quasiparticles lie close to the gap the energy loss during this process is $2\Delta$ and therefore the rate  is given by 

\begin{eqnarray}
 \Gamma_{N\rightarrow N,N_S+2\rightarrow N_S}=\frac{\dot Q_\mathrm{rec}(N_S+2)}{2\Delta}.
\end{eqnarray}

In these terms the heat balance then is fully described by the master equation for $P(N,N_S)$.